\documentclass[]{interact}

\usepackage{comment}
\usepackage{mwe}
\usepackage{tabularx}
\usepackage{multirow}
\usepackage{caption}
\usepackage{soul}
\usepackage{url}
\usepackage{apacite}

\begin{document}

% \articletype{ARTICLE}

\title{Evaluating User Experience and Data Quality in Gamified Data Collection for Appearance-Based Gaze Estimation}

\author{
\name{Mingtao Yue$^{a,\dagger}$, Tomomi Sayuda$^{a,\dagger}$, Miles Pennington$^{a}$, Yusuke Sugano$^{a}$}
\affil{$^{a}$Institute of Industrial Science, The University of Tokyo, Tokyo, Japan.\\$^{\dagger}$Both authors contributed equally to this research.}
}

\thanks{This is an original manuscript of an article published by Taylor \& Francis in International Journal of Human–Computer Interaction, available at: \protect\url{https://doi.org/10.1080/10447318.2024.2399873}. \\ CONTACT Yusuke Sugano. Email: \protect\url{sugano@iis.u-tokyo.ac.jp}}

\maketitle

\begin{abstract}
Appearance-based gaze estimation, which uses only a regular camera to estimate human gaze, is important in various application fields.
While the technique faces data bias issues, data collection protocol is often demanding, and collecting data from a wide range of participants is difficult.
It is an important challenge to design opportunities that allow a diverse range of people to participate while ensuring the quality of the training data.
To tackle this challenge, we introduce a novel gamified approach for collecting training data. 
In this game, two players communicate words via eye gaze through a transparent letter board. 
Images captured during gameplay serve as valuable training data for gaze estimation models. 
The game is designed as a physical installation that involves communication between players, and it is expected to attract the interest of diverse participants.
We assess the game's significance on data quality and user experience through a comparative user study.
\end{abstract}


\section{Introduction}\label{sec:intro}

The increasing importance of artificial intelligence (AI) and machine learning (ML) technologies highlights the urgent demand for training data. 
High-quality data is vital for building better ML models, but collecting sufficient training data presents significant challenges. 
For most ML tasks, training data needs diversity to guarantee that models operate robustly and without prejudice. 
Nonetheless, gathering data from diverse participants brings about several obstacles, particularly in privacy-sensitive areas involving facial images or biometric information.
The data collection is often time-consuming and burdensome, further discouraging participant recruitment.

Appearance-based gaze estimation~\cite{cheng2021appearance,ghosh2021automatic}, the ML task of estimating gaze direction from face images, is a prime example of these data collection challenges. 
The human gaze is considered a vital sensing modality in various application fields such as usability studies, hands-free input, and user state estimation for context-aware computing~\cite{majaranta2014eye}.
Appearance-based gaze estimation allows for eye tracking with only an ordinary camera, and this opens up the potential for various applications, including gaze-based interaction~\cite{bace2020combining,huynh2021imon} and attention monitoring~\cite{sugano2016aggregaze,zhang2017everyday,chong2017detecting}.
Training accurate and robust gaze estimation models requires diverse face images associated with gaze direction labels. 
Nevertheless, the standard data collection protocol, which requires participants to look at the gaze target for an extended period as per instructions, is extremely tedious.
Moreover, the inherent sensitivity of facial data causes concerns about privacy and ethical issues. 
This makes large-scale voluntary participation less appealing to the general public. 
Addressing these obstacles to achieve accessible and ethical data collection is a crucial research task.

\begin{figure}
    \centering
    \includegraphics[width=1\textwidth]{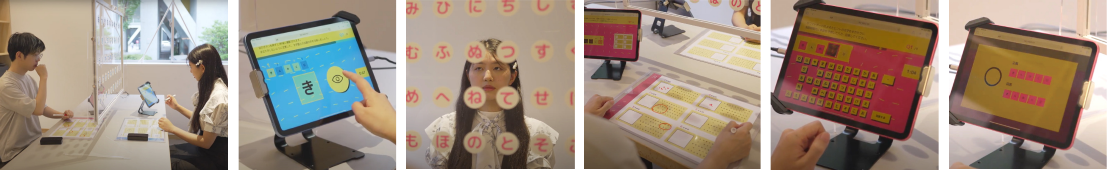}
    \caption{This paper proposes a mixed-media gamification system for collecting training data for gaze estimation.\label{fig:teaser}}
    % \Description{A collection of thumbnail images that give an appearance of the hardware and play of the gamified system.}
\end{figure}

This work addresses the challenges of collecting diverse and representative gaze data by introducing a novel gamified approach to appearance-based gaze estimation data collection. 
The primary objectives of this research are twofold: (1) to design a data collection system that facilitates the accurate recording of face images with corresponding gaze direction labels, and (2) to create an engaging and enjoyable gameplay experience that attracts a wide range of participants.
We hypothesize that the proposed gamified approach will yield training/evaluation data of comparable quality to standard data collection methods while providing a more engaging and motivating experience for participants. 
Although some previous examples exist of adding a gamification aspect to ML data collection processes~\cite{roemmele2014triangle,bragg2021asl}, there was no attempt where gamified data collection was applied to appearance-based gaze estimation.

As illustrated in Fig.~\ref{fig:teaser}, the game involves two players sitting opposite each other with a transparent letter board between them. 
Inspired by augmentative alternative communication techniques for people with motor disabilities~\cite{zeitlin1995use}, it is designed as a quiz-like game where players communicate words using their gaze.
One player is provided some letters of the word as hints, and the other player conveys the missing letters by using their gaze. 
This not only provides an enjoyable quiz experience but also enables us to collect training and evaluation data for gaze estimation by capturing facial images of players who use their gaze to convey letters.
We have designed this game to be a cooperative two-player game that incorporates physical elements and face-to-face interaction. 
By doing so, we aim to create an environment that is effective and appealing to a diverse audience for data collection.

In this paper, we report the results of user experiments conducted in a public cafeteria on a university campus. 
We obtained questionnaire responses from 46 out of 47 participants, allowing us to perform a subjective assessment of their gaming experience. 
Furthermore, a subset of 22 participants were asked to play the game while equipped with a wearable eye tracker. 
Using this data, we quantitatively evaluated gaze behavior during the game experience, i.e., how accurately players look at target letters. 
We also analyzed the characteristics of the data obtained from the game by comparing our results to outputs from pre-trained gaze estimation models. 
In this way, we validate that our game can acquire high-quality training data from a diverse range of participants.

To the best of our knowledge, this work represents the first attempt to apply gamification techniques to the problem of collecting appearance-based gaze estimation data. 
By bridging the gap between engaging user experiences and rigorous data acquisition, our approach opens up new possibilities for collecting diverse and representative gaze datasets in an accessible and enjoyable manner.
Potential use cases for a gamified approach include supplementing in-house data collection efforts, enabling data collection from hard-to-reach populations, and promoting public engagement with gaze estimation technology. 
The democratization of data collection through gamification is expected to offer an alternative approach to more equitable and inclusive AI training. 
At the same time, the trade-offs between data diversity and quality control must be weighed based on the specific application requirements. 
This study's insights regarding data quality are expected to offer implications for practical deployments in this regard. 

The main contributions of this work are as follows:
\begin{itemize}
    \item The development of a novel gamified data collection system for appearance-based gaze estimation that combines engaging gameplay with accurate gaze data acquisition.
    \item A real-world user study comparing the gaze label accuracy and participant experience between the proposed gamified approach and a standard data collection method.
    \item Insights into the characteristics and potential utility of the collected gamified dataset for training and evaluating gaze estimation models.
    \item Design recommendations and lessons learned for implementing effective and engaging gamified data collection systems in the context of appearance-based gaze estimation.
\end{itemize}

\section{Related Work}

\subsection{Appearance-based Gaze Estimation and Datasets}

Conventional remote eye tracking devices, utilizing specialized hardware, have substantial restrictions on the range of head movement of the target person~\cite{hansen2009eye}. 
In contrast, gaze estimation techniques using regular cameras and computer vision technology have the potential to resolve these limitations, enabling gaze measurement in more natural settings. 
With the advancement of deep learning, appearance-based gaze estimation has been actively studied in the last decade~\cite{cheng2021appearance,ghosh2021automatic}. 
Appearance-based gaze estimation is an approach that builds an ML-based model from large-scale data, and the model can output gaze direction from an input face image without explicitly extracting image features such as the center of the pupil. 
Such an approach is known to be robust even with low-resolution and poor-quality images~\cite{zhang2019evaluation}, and there is potential for even more diverse eye tracking applications~\cite{sugano2016aggregaze, zhang2017everyday, bace2020combining}.
When taking an appearance-based approach, the quality and diversity of training data pose significant challenges.

A classical approach to appearance-based gaze estimation has assumed data specificity to an individual~\cite{tan2002appearance,lu2014adaptive,sugano2012appearance}. 
Recent years have witnessed an increased interest in training models that are not person-dependent, utilizing large-scale data collected from various individuals.
Deep learning models have proved highly effective in this endeavor, and various gaze estimation models have been proposed in the literature~\cite{zhang2017s, park2018deep, cheng2022gaze, park2019few}.
Similarly to other image recognition tasks, appearance-based gaze estimation also suffers from overfitting to training data, making accurate estimation in testing environments that significantly differ from the training data quite challenging. 
Such issues cannot be resolved solely through domain adaptation or generalization approaches~\cite{cheng2022puregaze,jindal2023contrastive, wang2022contrastive, lee2022latentgaze, qin2023domain}, and diverse datasets are indispensable both for the verification of robustness and for the training of more powerful models.

Several datasets have been created to date to ensure the diversity of various factors such as the lighting environment and head pose~\cite{sugano2014learning,funes2014eyediap,krafka2016eye,zhang2017mpiigaze,fischer2018rt,kellnhofer2019gaze360,zhang2020eth, smith2013gaze}.
Such datasets can be constructed by instructing participants to look at specific target positions and repeatedly capturing images of their faces at the moment of fixation. 
For instance, in the Gaze360 dataset, participants stand around a 360-degree camera while researchers move a gaze target board marked with an AR marker.
In the ETH-XGaze dataset~\cite{zhang2020eth}, participants seated in front of a screen were instructed to click the mouse to confirm when a circle displayed as a gaze target shrinks to a dot.
However, the absolute number of these studies is relatively small compared to research proposing methods for gaze estimation. 
Inevitably, there is a bias in the locations where data collection is performed and in the attributes of the participants. 
Approaches to synthesizing training data have also been explored, but their performance is still inferior to real data~\cite{qin2022learning, wood2016learning, wood20163d}.
Furthermore, as mentioned earlier, appearance-based gaze estimation requires facial images, which introduces fundamental difficulties in recruiting a wide range of participants. 
This work constitutes a novel attempt to address the challenge of collecting diverse data through gamification to overcome the prevailing difficulties in this field.

\subsection{Games for ML Data Collection}

Gamification reaches a larger audience by altering a subject matter to be experienced as a game~\cite{seaborn2015gamification} and has been investigated since the early 1980s~\cite{malone1980what}.
The concept of gamification has been discussed across various fields of study, but it is particularly prominent within the context of crowdsourcing~\cite{morschheuser2017gamified}. 
The idea typically revolves around distributing gamified tasks to users to utilize human knowledge to solve larger tasks such as protein structure discovery~\cite{cooper2010predicting, khatib2011algorithm}. 
However, such an approach does not always align with ML data collection, which is the focus of our research. 
Specific instances where the main focus is collecting data annotations include image labels~\cite{von2006improving, von2004labeling}, common-sense texts~\cite{von2006verbosity}, image preferences~\cite{hacker2009matchin}, and sound/music annotations~\cite{law2007tagatune, turnbull2007game}.
In most of these cases, the scenario envisioned is one wherein numerous players on the web collectively perform annotations for the data. 
Our research's objective is to design a game that allows for the simultaneous acquisition of the data (in our case, face images) and the corresponding annotations (gaze directions). 

As more closely related work, Roemmele et al. proposed a game to collect training data for motion-based action recognition~\cite{roemmele2014triangle}.
The paper presents Triangle Charades, a web-based online game where players animate triangles to depict actions for others to guess.
Bragg et al. also proposed a mobile game called ASL Sea Battle to collect a large corpus of labeled American sign language videos in an entertaining way~\cite{bragg2021asl}.
Players record videos of themselves signing to attack squares on a grid labeled with signs, and their opponent views the videos to label them and reveal if it was a hit on their hidden ships.
Compared to these cases, our study has two major uniquenesses.
One is that this study's target for data collection is the training data for appearance-based gaze estimation, which significantly differentiates our research from the above examples. 
Another unique aspect of our study is that our game design is not purely an online application. It is designed as a mixed-media system that incorporates physical devices. 

Exploration of gamification elements is also seen in the field of citizen science~\cite{bowser2013using, lintott2008galaxy}. 
Regarding sparking interest and understanding of technology, our motivations are similar to those of citizen science. 
However, the unique aspect of our study that sets it apart, even in this context, is gaze estimation targeting through a mixed-media game. 
While physical design is necessary for our purposes, it is not obvious that this will work in practice.
This study evaluates the user experience and data collected through a real-world user study.

\section{Proposed System}

\begin{figure}[t]
    \centering
    \includegraphics [width=1.0\linewidth]{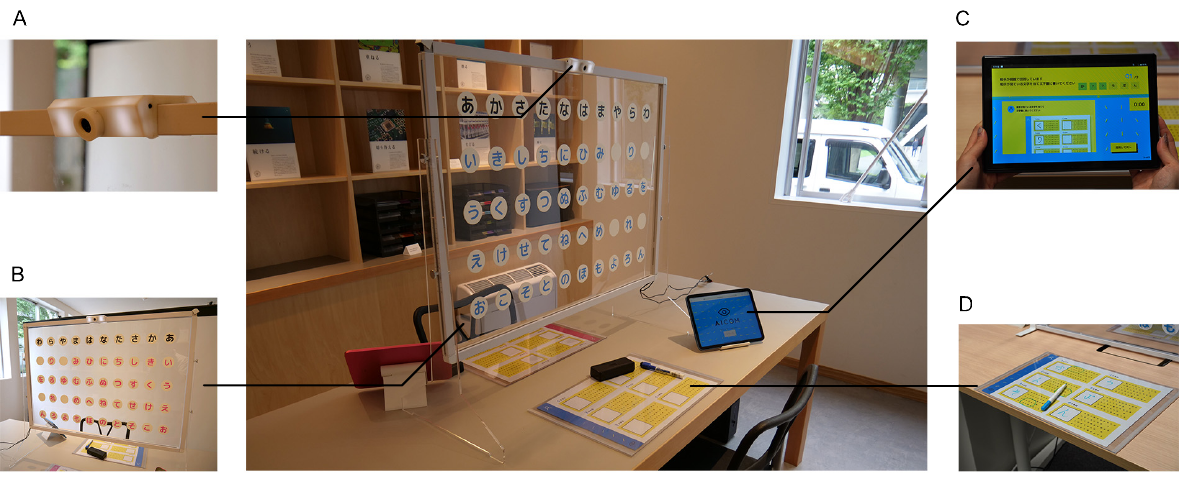}
    \caption{Hardware overview of the gamified data collection system. Our system consists of A) two webcams, B) a transparent letter board, C) tablet PCs, and D) custom writing boards.\label{fig:hardware}}
    % \Description{Photographs showing the game hardware used in the experiment; a close-up photograph accompanies each component from A to D.}
\end{figure}

In this section, we will discuss the details of the design and implementation of our gamified data collection system.
Fig.~\ref{fig:hardware} shows an overview of our game system.
Our system consists of a transparent letter board (Fig.~\ref{fig:hardware} B) equipped with two webcams embedded in the frame (Fig.~\ref{fig:hardware} A), tablet PCs (Fig.~\ref{fig:hardware} C), and custom writing boards (Fig.~\ref{fig:hardware} D).
The webcams are physically connected to a server PC, and the tablet PCs establish a connection with this server using a local wireless LAN via a browser. 
The two participants engage in the game while seated opposite one another, separated by the letter board. 
Japanese alphabet (Hiragana) is displayed on the letter board, reflecting the language native to the study's location. 
The letters are organized to appear reversed when seen from the opposing player's side. 
Consequently, from either player's perspective, specific physical positions on the board coincide with the same letters.

\subsection{Design Process}

As previously mentioned, the design process has two main objectives. 
The first objective is to facilitate recording face images with as precise gaze direction labels as possible, which are used for training appearance-based gaze estimation models. 
The second objective is to design a gameplay experience that allows the general participants to engage with the technology and ensures they enjoy the process. 
The game aims to appeal to a broad audience and facilitate the collection of diverse and representative data by ensuring accurate data collection while providing an engaging and accessible experience.

\subsubsection*{Cooperative Gameplay}

The first objective of acquiring accurate gaze direction labels is mainly reflected in the game logic design with a cooperative two-player setup. 
The training data for appearance-based gaze estimation must contain facial images linked to their ground-truth gaze positions.
In our game, the questioner player is incentivized to convey the letters precisely to the answerer player through eye gaze, as the success of both players depends on the accuracy of the information conveyed. 
This design decision addresses the challenge of ensuring players focus on the designated targets since merely instructing them does not guarantee compliance.

Both players share a common goal and must work together to succeed in a cooperative setup. 
This design choice encourages the questioner to focus on the target letters accurately, as their partner's success depends on receiving correct information. 
In contrast, a competitive design may motivate players to deliberately mislead their opponents by looking at different letters, compromising the quality of the collected data. 
The cooperative design promotes a more engaging and enjoyable experience, as players must collaborate and communicate effectively to achieve their shared objective.
Such a cooperative design focused on accurately conveying messages to other human players has also been adopted in the prior work~\cite{roemmele2014triangle}.

\subsubsection*{Physical Game Design}

To address the second objective of creating an engaging gameplay experience, we incorporated physical elements and face-to-face interaction between players.
The face-to-face setup facilitates non-verbal communication and social interaction, creating a more immersive and enjoyable experience than a digital interface.
Furthermore, it should be noted that the task of interpreting the gaze of another cannot be achieved without employing a design that physically incorporates a transparent letter board.

The two-player setup also allows for collecting gaze data in a more natural and interactive context than the standard data collection. 
In the standard data collection protocol, participants are typically seated alone and instructed to fixate on specific targets, which may feel repetitive and unengaging. 
The gamified approach introduces a social element, where players must interpret and respond to their partner's gaze cues, simulating a more realistic scenario for nonverbal interaction. 
This dynamic may elicit more genuine and diverse facial expressions that better reflect real-world use cases.

Furthermore, the physical setup of the game, with players seated opposite each other, is designed to attract the attention of passersby in a public setting. 
Using a transparent letter board and writing boards adds a tangible dimension to the game, making it more accessible and appealing to a diverse audience. 
This design aims to encourage participation from diverse individuals and collect data from a variety of populations.
As pointed out in prior research aimed at certain demographics, such as children and the elderly~\cite{xie2008tangibles,bakker2015peripheral,garcia2017tangibot,yuan2021tabletop}, the advantages of intuitive and communicative physical device designs can be important when creating tools accessible to a diverse user base.

\subsection{Gameplay Flow}

\begin{figure}[t]
    \centering
    \includegraphics [width=\linewidth]{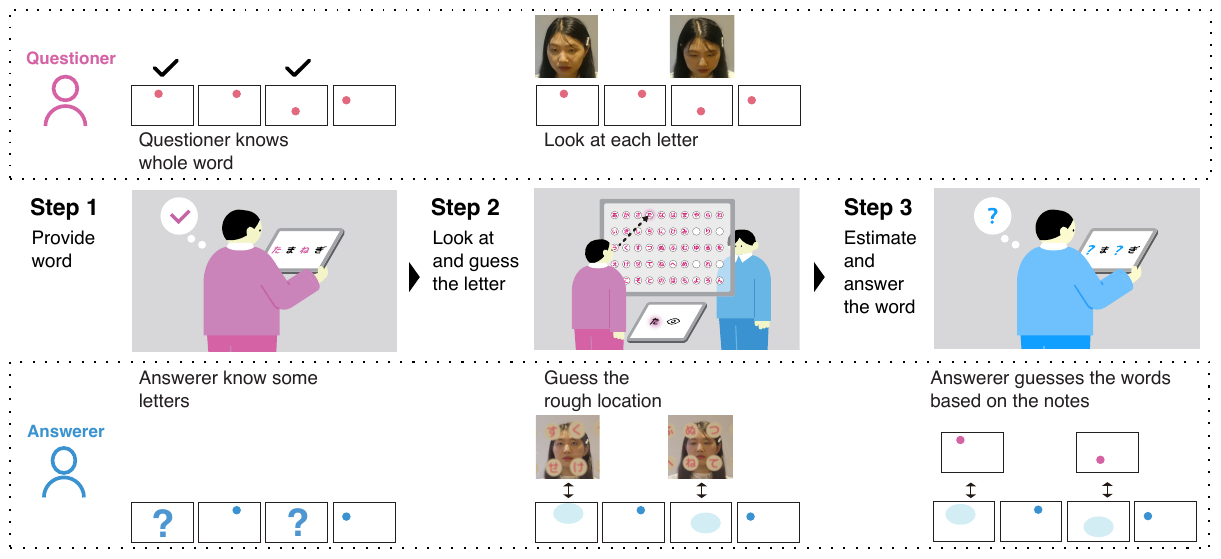}
    \caption{The flow of the gameplay. Step 1: The questioner first identifies the word, and the answerer observes some clue letters. Step 2: The questioner looks at each letter, and the answerer guesses the rough locations of each letter. Step 3: The answerer guesses the word.\label{fig:experience}}
    % \Description{An illustration explaining the concept and basic rules of the game.}
\end{figure}

Figure~\ref{fig:experience} presents the process of gameplay. 
Initially, the system prompts the questioner with the word for the quiz. 
At this stage, a few letters from the same word are disclosed to the answerer to function as hints. 
In the subsequent second step, the questioner communicates the hidden letters to the answerer using eye gaze, not verbal cues, by looking at each letter on a transparent board. 
The answerer does not need to specify the letter but rather estimate its approximate position by interpreting the questioner's rough gaze direction.
Customized writing boards, which show the same letter layout as the transparent board, aid this process by allowing the answerer to draw a circle around the expected rough letter location based on their interpretation of the questioner's gaze direction.
This step is executed repeatedly for each hidden letter. 
Finally, in the third step, the answerer guesses the word, identifying the missing letters based on the estimated positions to form a coherent word. 
If the word is guessed correctly, the duo scores a point. 
The players exchange roles, and the game proceeds with the next word.
Through the game, the questioner's face images and the positions of the hidden letters are saved. 
This is equivalent to capturing face images with known gaze target locations, providing data that can be used for training appearance-based gaze estimation models.

Our game is designed for two-player participation, alternating roles between questioners and answerers. 
The questioner's objective is to indicate a word presented by the system to the answerer by gazing successively at letters on a transparent board. 
Since identifying the exact letter another player looks at is more challenging than one might assume, we have eliminated the need for the answerer to identify each letter individually. 
Instead, the answerer focuses on the entire set of letters, attempting to discern the complete word to ensure it forms a sensible term. 

Additionally, we observed that guessing a word with no clues is extraordinarily demanding. 
During the prototyping phase, we realized that unveiling certain letters of a word as hints significantly mediates the difficulty level. 
Maintaining this level, neither too simple nor overly complicated, is crucial to sustaining the players' interest in the game. 
Our strategy benefits from the fact that the game's difficulty can be adjusted by manipulating the number of letters presented as hints.

\subsection{User Interface}

\begin{figure}[t]
    \centering
    \includegraphics [width=1.0\linewidth]{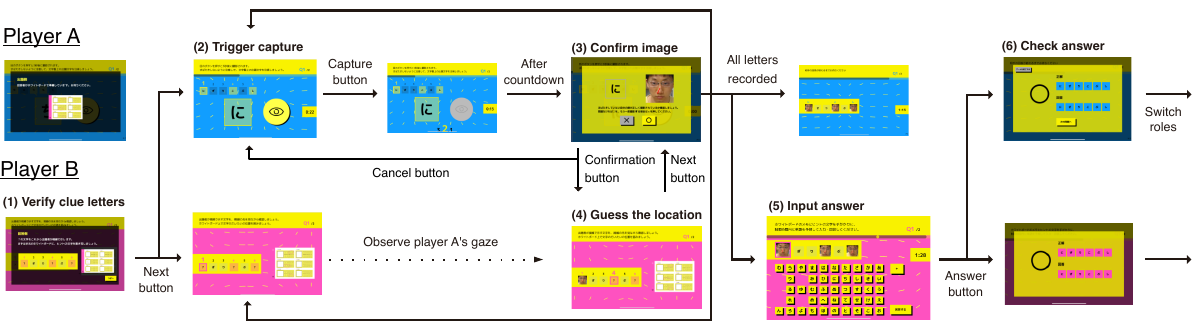}
    \caption{Flow of our user interface. (1) The questioner and the answerer confirm the target letter, and (2) the questioner triggers the image capture. (3) Once the questioner has approved the captured image, (4) the answerer anticipates and notes the rough position of the letter. (5) When all letters have been captured, the answerer guesses the word and inputs the answer. (6) The system displays the correct answer, and players switch roles and move on to the next word.\label{fig:UI-screen}}
    % \Description{Screen transition diagram of the implemented user interface. The screens of both players and their transition destinations are shown.}
\end{figure}

Figure~\ref{fig:UI-screen} presents the flow of our implemented user interface. 
The game starts when both players hit the start button. 
Before the first question, a brief game description and notes appear on both screens.
At first, player B, the answerer, gets a prompt indicating which letter will be conveyed and can verify the clue letters displayed on their screen.
The game enters the phase of conveying letters when the answerer gets ready and presses the next button.

The questioner fixates on the location of each hidden letter on the transparent board and presses the capture button. 
After the countdown, the system captures and displays an image of the questioner's face, combined with gaze estimation results computed by a pre-trained gaze estimation model. 
When no face is detected in the captured image, the system defaults to displaying a text message rather than a face image. 
The questioner presses the confirmation button after confirming that their face has been accurately captured without blinking. 
The system then presents the captured face image to the answerer. 
Concurrently, the answerer endeavors to interpret the questioner's gaze and anticipate the probable location of the letter on the writing board. 
The game progresses to the subsequent letter once the answerer completes their prediction and presses the next button.

Upon the questioner completes all the letters, the system shifts focus to the answerer, awaiting their response. 
Subsequently, the answerer inputs their predicted words using the on-screen keyboard. 
The captured face images associated with each hidden letter are also exhibited side by side.
The screen also shows a timer indicating the elapsed time on the top, and the answer must be provided within a preset duration. 
As this time limit draws near, an additional clue is revealed in the form of the first letter. 
The system shows whether the answer is correct once the time limit concludes or the answerer finalizes the answer. 
Moving to the next word, the players switch roles; this change is initiated by the answerer pressing the proceed button. 
The gaming session ends after a specific quantity of question words has been presented.

\subsection{Implementation Details}

The positional relationship between the letter board and the camera is calculated by attaching a camera calibration pattern to the board and observing it through a mirror~\cite{rodrigues2010camera}.
This process determines the 3D position of each letter in the camera coordinate system, which is essential for accurately mapping the letters to the corresponding 3D gaze directions.

The frontend of the system is developed using React and Node.js.
The backend server, implemented using Python and the Flask framework, handles the game logic, data storage, and processing. 
The frontend communicates with the backend server through a WebSocket connection, enabling real-time data exchange and synchronization between the two components. 
The backend maintains a database of word dictionaries, which are curated from a Japanese language learners' dictionary. 
The server selects words from the dictionary based on the game settings, such as the number of question words per game, minimum and maximum letters per word, and the number of hidden (question) letters in each word. 
Note that special characters not used on the letter board, such as letters with diacritics, are excluded from the questions.
If the question word contains these characters, they will always be displayed as hints.
The backend also controls the game's time limit and the timespan until extra clue letters are revealed. 

In addition to the word selection, the backend server also handles the image capture and preprocessing. 
The server controls the webcams attached to the letter board and captures the questioner's face images at the appropriate moments during gameplay. 
The captured images are then processed through face detection~\cite{bazarevsky2019blazeface} and facial landmark alignment~\cite{bulat2017far}. 
Then, the face images are normalized following the standard protocol in appearance-based gaze estimation~\cite{zhang2018revisiting} to ensure consistency and compatibility with existing gaze estimation models.
To provide real-time feedback to the players, the backend server also performs gaze estimation on the preprocessed face images using a pre-trained model. 
The gaze estimation model used in this implementation is pre-trained on the MPII-NV dataset~\cite{qin2022learning}. 
The estimated gaze direction is visualized on the frontend interface.

\section{User Study}

\begin{figure}[t]
    \centering
    \includegraphics [width=1.0\linewidth]{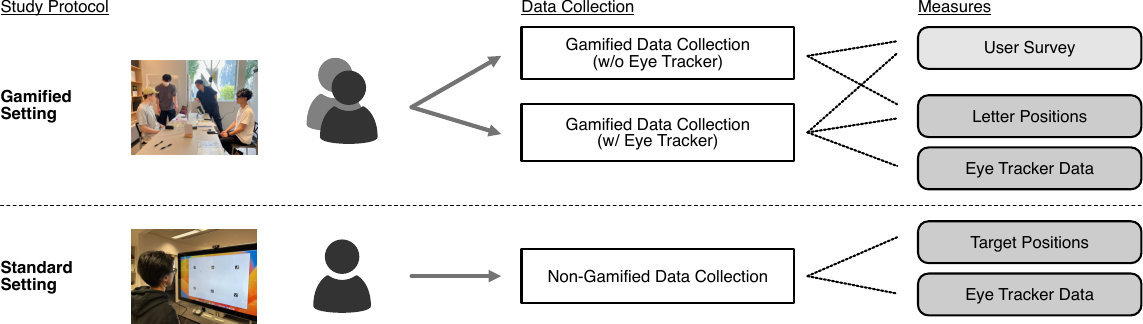}
    \caption{Illustration of the user study settings. The gamified setting was conducted in a public space, where voluntary participants were visitors. The standard-setting experiment was conducted in our laboratory, and participants were recruited directly. We recorded eye tracking data from gamified and standard settings for label accuracy comparison and asked participants to complete a custom survey.\label{fig:study}}
    % \Description{Flowcharts illustrate the procedure for the two experimental conditions. In the voluntary participation gamified setting, participants engage in gamified data collection with or without an eye tracker. Their letter positions are recorded, and they answer the user survey. If they participate with an eye tracker, their data is also recorded. In the standard setting, where subjects are directly recruited, they collect non-gamified data and log both target positions and eye tracker data.}
\end{figure}

We conducted a user study to compare the accuracy of the gaze label acquired under the gamified and standard settings. 
Fig.~\ref{fig:study} illustrates the flow of the two study conditions.
The first gamified setting involved experiments conducted in a public space, where we made a gamified system available for cafeteria visitors to experience.
The second standard setting reflected the conventional data collection method within a controlled laboratory environment, where we explicitly recruited participants and provided them with specific instructions.

We employed wearable eye trackers to obtain the reference gaze positions and investigated the differences in gaze label accuracy between the two settings. 
Since wearing an eye tracker makes facial images unsuitable for model training and evaluation, only a subset of participants were equipped with eye trackers in the gamified setting.
Additionally, we evaluated user experience using a custom survey under the gamified setting.
The following subsections provided further details of the study protocols and additional measures.

\subsection{Study Protocols}

\subsubsection{Gamified Setting}

% \begin{subfigure}[b]{0.35\textwidth}
%     \centering
%     \includegraphics[width=\textwidth]{fig/experiment.pdf}
%     \caption{System installation\label{fig:experiment}}
%     % \Description{A group of images of the experiment. Various participants are experiencing the game together with our staff.}
% \end{subfigure}
% \hfill
% \begin{subfigure}[b]{0.6\textwidth}
%     \centering
%     \includegraphics[width=\textwidth]{fig/eyetracker.pdf}
%     \caption{AR markers and eye tracking headset\label{fig:eyetracker}}
%     % \Description{Photos showing the wearable eye tracker used in the experiment and the AR markers used to map the measurement results to the letter board.}
% \end{subfigure}

\begin{figure}[t]
    \centering
    \includegraphics [width=1.0\linewidth]{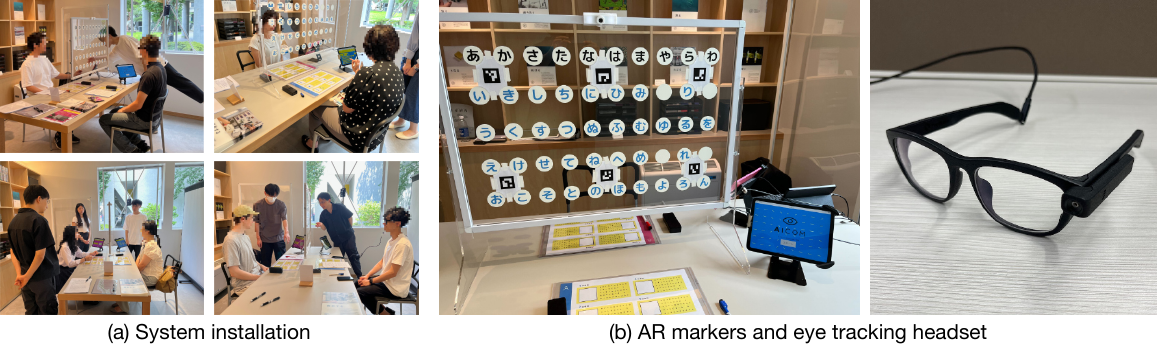}
    \caption{The environment of our gamified data collection study. (a)~We installed the game system in the corner of a cafeteria on our university campus. (b)~We collected eye tracking data from a subset of participants while playing the game.\label{fig:user_study}}
\end{figure}

In keeping with the concept of our system, we did not recruit experiment participants individually for the gamified data collection. 
Instead, we installed our system in a cafeteria on our university campus for five days and invited volunteers to participate. 
This gave us an ideal setting to conduct our experiments with unconstrained participants. 
To this end, we installed the game system in the corner of the cafeteria (Fig.~\ref{fig:user_study}(a)). 
Even though this cafeteria is located on a university campus, it is open to university affiliates and the general public. 
Considering privacy concerns, we placed a large screen behind the system's non-wall-facing side, which acted as a shield to prevent accidentally capturing bystanders in the background. 
At the cafeteria entrance, we put up posters announcing that an experiment was being conducted. 
Likewise, we placed similar posters near the experiment booth, and another poster was set up to remind individuals that filming was underway. 
These measures ensured that all visitors knew about the ongoing experiment and the recording activities.

We ensured that at least two staff members were always at the experiment booth. 
If a person showed interest in our experiment, they were first handed a detailed handout explaining the purpose and scope of the experiment and conducted through a thorough explanation concerning the data collection process. 
At this time, it was explicitly explained that face images would be captured and that they could be used for non-commercial research as a public dataset\footnote{The dataset is being considered for publication in a future long-term study and will not be published with this paper.}. 
We allowed only those who expressed understanding and consent to the data collection process and voluntarily signed an agreement form to participate in the experiment. 
Since our game requires Japanese vocabulary, the participants were Japanese speakers, and the explanation and confirmation of consent were also conducted in Japanese.
The university's ethics review committee examined and approved the experiment details and consent process.

Although such cases were rare, individuals who came alone showed interest in participating in the experiment.
Occasionally, two strangers who visited simultaneously were paired to participate in the experiment.
If a partner could not be found, the participants were paired with a staff member for the experience.
In this case, the data obtained from the staff member were not used in the following analysis.
To ensure that all participants could experience the roles of questioner and respondent, we arranged for everyone to engage in two game rounds. 
Throughout the experiment, we standardized the word length to five or six characters, with the number of characters in question consistently set at three. 
If any participant encountered difficulties understanding how to progress in the game, our staff could provide immediate assistance. 

Over five days of system installation, we collected data from 47 participants.
We assigned each participant to produce data for three letters in their role as a questioner. 
However, due to a system glitch, data from one participant was only recorded for two letters, resulting in a total of data gathered for 140 letters.

\subsubsection{Standard Setting}

The standard setting refers to an experiment conducted in a situation similar to the usual data collection performed in previous studies.
The experiment was conducted within our laboratory space, and we secured participants by directly approaching students and researchers.
Therefore, it is assumed that participants in the standard setting engaged in the experiment with a more serious attitude compared to those in the gamified setting.

We showed a white canvas with the same physical size as the letter board on a 55-inch display.
The participants were instructed to look at 50 stimuli randomly shown inside the canvas.
We showed a circle of the same size as the letter board design (Fig.~\ref{fig:hardware}) and a plus sign at the center of the circle.
Whenever the participant pressed the space key of a keyboard, a circle began to shrink to the center. 
After three seconds, the same period as the capture countdown for the gamified setting, the shrinking circle disappeared, and gaze data and scene images were retrieved from the eye tracker.

Sixteen participants (thirteen male and three female) took part in this non-gamified standard data collection.
The ages ranged from 22 to 30 years ($M=25.5$, $SD=2.5$). 

\subsection{Measures}

\subsubsection{Gaze Label Accuracy}

Under both the gamified and standard settings, we used a wearable eye tracker to simultaneously capture the ground-truth gaze positions to evaluate the accuracy of the acquired gaze labels.
In other words, for each data point, we obtain two gaze directions: one calculated from the character or gaze target that the participant was looking at during the game or data collection, and the other measured by the eye tracker. 
As it is possible for participants to unintentionally look at different places during the game or data collection, their facial images could be captured with their gaze directed elsewhere. 
This experiment aims to verify how closely the gaze labels obtained under the two settings approximate the ideal ground truth labels captured by the eye tracker.

In a gamified setting, eye tracking data was collected from only a portion of the participants, specifically 22 out of 47.
As mentioned earlier, this is because the distinct appearance of the eye tracker frame acts as an artifact on the facial image, making the captured face images unsuitable for training and evaluation data. 
Each participant was instructed to wear the eye tracker as the questioner while playing the game.
We used the Pupil Invisible wearable eye tracker\footnote{\url{https://pupil-labs.com/products/invisible/}} to collect eye tracking data.
Before each game session, we conducted gaze offset correction using the Pupil companion app.
We instructed the participants to look at the upper left corner of each AR marker and adjusted the offset between the marker corner and gaze location.
For these game sessions, the backend server accessed the real-time API of Pupil Invisible to retrieve eye tracking data at exactly the moment when the questioner's face image was captured.
Furthermore, we collected eye tracking data from all 16 participants under the standard setting. 
We also performed a similar offset correction for each participant before collecting the data.
Eight data points were excluded due to severe degradation of scene image quality, resulting in 792 data points.

To map the eye tracking data to the board coordinate system, we placed six ArUco AR markers~\cite{garrido2014automatic} on both sides of the board before the game sessions (Fig.~\ref{fig:user_study}(b)). 
In the standard setting, we also displayed AR markers at the same positions on the display.
After retrieving the data from Pupil Invisible, AR markers were detected in the scene image using the OpenCV library~\cite{bradski2000opencv}. 
If the number of detected markers was less than two for each row, the data was handled by manually annotating more marker locations.
Then, by leveraging the intrinsic parameters of the scene camera provided by Pupil Invisible and the marker locations in the board coordinate system, we obtained a homography matrix to transform the image coordinate system into the board coordinate system.
In addition, we estimated the camera origin location using the PnP algorithm~\cite{lowe1991fitting}.
This way, we compute the angular distance between the gaze and letter positions, assuming that both vectors originate from the camera origin.

\subsubsection{User Experience and Technology Acceptance}

After the game experience, participants in the gamified setting were encouraged to provide feedback through an online questionnaire. 
Many of the experiment participants were ordinary people who happened to visit the cafeteria and had limited time to respond to the survey after the experiment.
In addition to gameplay, we were also interested in additional questions about how participants' perceptions of gaze estimation and AI/ML might shift following their game experience. 
For these reasons, we designed a short custom survey with items specifically focused on gameplay and technology acceptance. 
Please note that the standard setting lacks gaming elements, and we did not ask for any subjective feedback in this setting.

In the questionnaire, besides gathering information about the participant's gender, age, and industry, we asked them to respond to the statements listed in Tab.~\ref{tab:qq} on a five-point Likert scale ranging from ``\textit{strongly agree}'' to ``\textit{strongly disagree}''.
Moreover, we asked for feedback in a free-form format, allowing participants to express their thoughts openly and without restriction. 
We obtained survey responses from 46 out of 47 participants, comprising 26 males and 20 females. 
The participants' ages ranged from 18 to 79 years ($M=37.7$, $SD=14.8$).

\begin{table}[t]
\caption{List of quantitative questions in our survey. Participants were instructed to provide five-point Likert scale ratings.\label{tab:qq}}
\small
\begin{tabularx}{\linewidth}{llX}
    \toprule
    \textbf{Category} & \textbf{ID} & \textbf{Question}  \\
    \midrule
    \multirow{5}{*}{Game} 
    & Q1-1 & It was interesting to read the gaze of the other player.  \\
    & Q1-2 & It was easy to read the gaze of the other player. \\
    & Q1-3 & It was interesting to guess the whole word. \\
    & Q1-4 & It was easy to guess the whole word. \\
    & Q1-5 & It was interesting to cooperate with the other player. \\
    & Q1-6 & I would like to play the game again. \\
    \midrule
    \multirow{4}{*}{Technology}
    & Q2-1 & The game was a good opportunity to learn about gaze estimation technology. \\
    & Q2-2 & I could understand how the game could lead to research on gaze estimation AI. \\
    & Q2-3 & I could understand why collecting face images of various people through the game is necessary. \\
    & Q2-4 & I am not afraid that my face image will be used for AI and ML research. \\
    \bottomrule
\end{tabularx}
\end{table}
\section{Results}

\subsection{Data Accuracy Compared to Standard Data Collection}\label{sec:eyetracker}

We first compare the accuracy of gaze labels under the two settings using data acquired with the eye tracker.
As mentioned above, we used 65 data points collected from 22 participants in the gamified setting and 792 from 16 participants in the standard setting.
Please note that individual differences are expected to play a larger role in the accuracy of the gaze label.
While there are more data points for the standard setting, the number of participants is approximately in the same order for both settings.

\begin{figure}[t]
    \centering
    \includegraphics [width=1.0\linewidth]{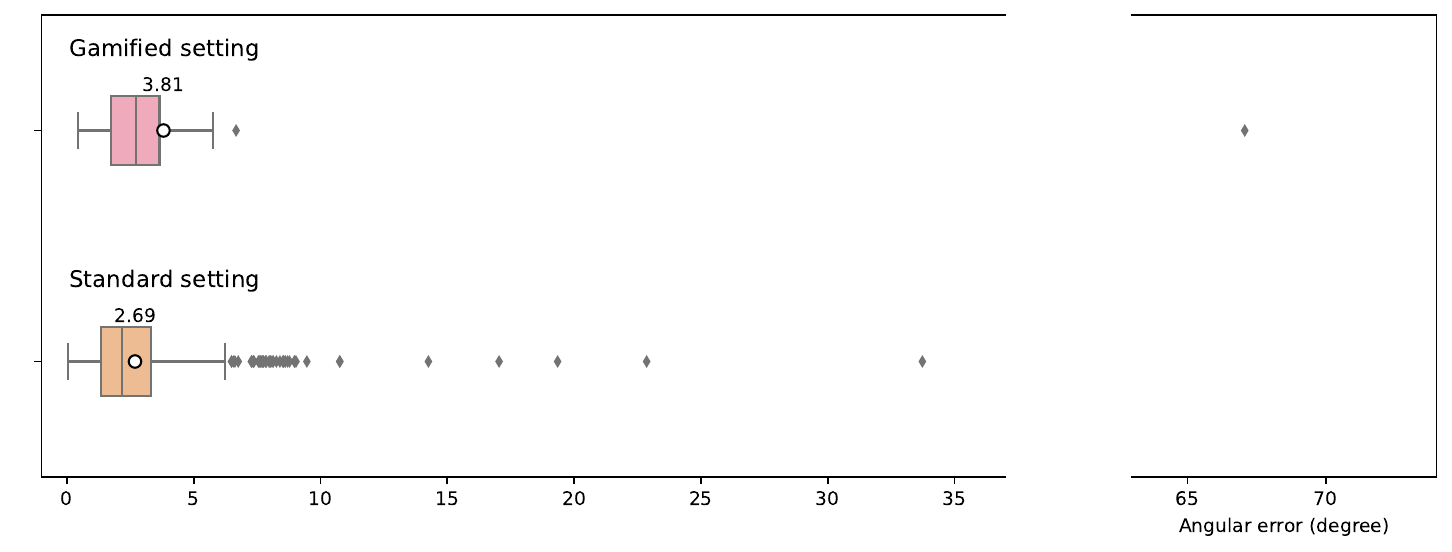}
    \caption{Angular error distributions of two data collection settings. Each box plot corresponds to gamified and standard data collection, and the horizontal axis shows angular error evaluated from the eye tracking data.\label{fig:angular_comparison}}
    % \Description{Two box plots corresponding to the gamified and standard data collection settings.}
\end{figure}

Figure~\ref{fig:angular_comparison} presents box plots illustrating the distribution of errors among participants in the gamified and standard settings. 
Because of an extreme outlier, the horizontal axis is partially omitted in the middle for better visibility.
In the gamified setting, represented as a plot at the top, the mean angular error was indicated as $3.81$ degrees. 
Meanwhile, the standard setting at the bottom reflected a mean angular error measuring $2.69$ degrees. 
Statistically, the gamified setting had significantly higher errors than the standard setting (Mann-Whitney U test, $U=30318$, $p=0.02$).
One evident outlier in the data from the gamified setting was captured when the participant was looking not at the letter but at the tablet. 
After manually eliminating this outlier, the mean angular error of the gamified setting was $2.82$ degrees, which is comparable to the standard setting.

\begin{figure}[t]
    \centering
    \includegraphics [width=1.0\linewidth]{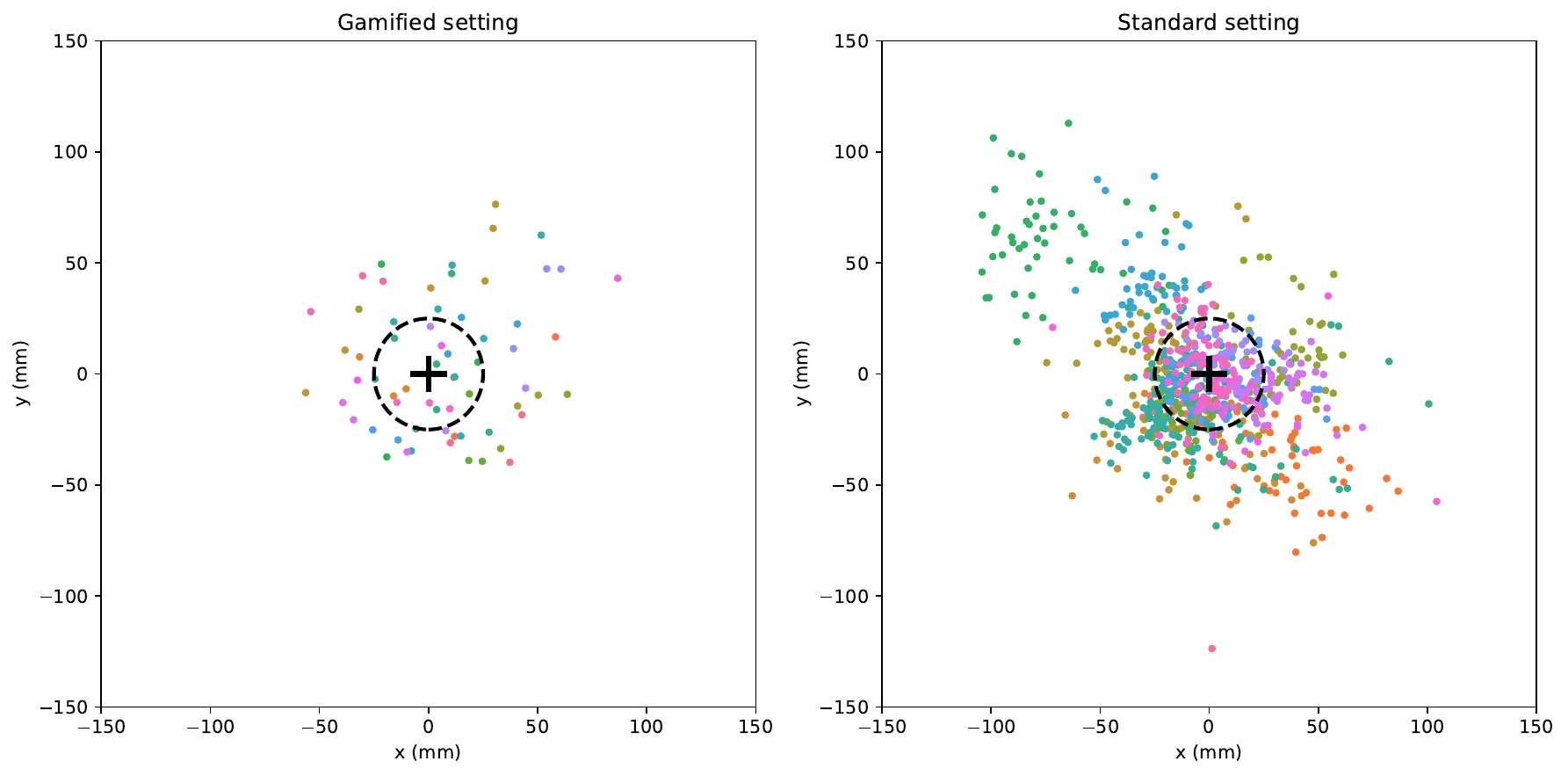}
    \caption{Spatial distributions of measured gaze positions. The center of each plot corresponds to the expected gaze target locations and shows scattered plots of the relative gaze positions measured by the eye tracker. Each data point is color-coded by the participant's ID.\label{fig:relative_position}}
    % \Description{Two scatter plots of measured gaze positions. In the center is a cross mark and a dotted circle around it, on which dots of different colors are distributed.}
\end{figure}

Figure~\ref{fig:relative_position} visualizes the distribution of gaze positions for both gamified and standard settings. 
We plotted the relative measured gaze position from the assumed target location for both scenarios. 
Each data point is color-coded by the participant's ID, and we have also added circles of the same size as on the text version for better representation. 
The visualization range is limited to $\pm 150$ mm, but please note that outliers (one for gamified setting and five for standard setting) extend beyond this range as in Fig.~\ref{fig:angular_comparison}. 
While a slight spatial bias can be observed, it is apparent that the proposed approach has not resulted in any extreme skewing of the distribution. 

\subsection{Analysis using a Pre-trained Gaze Estimation Model}

\begin{figure}[t]
    \centering
    \includegraphics [width=1.0\linewidth]{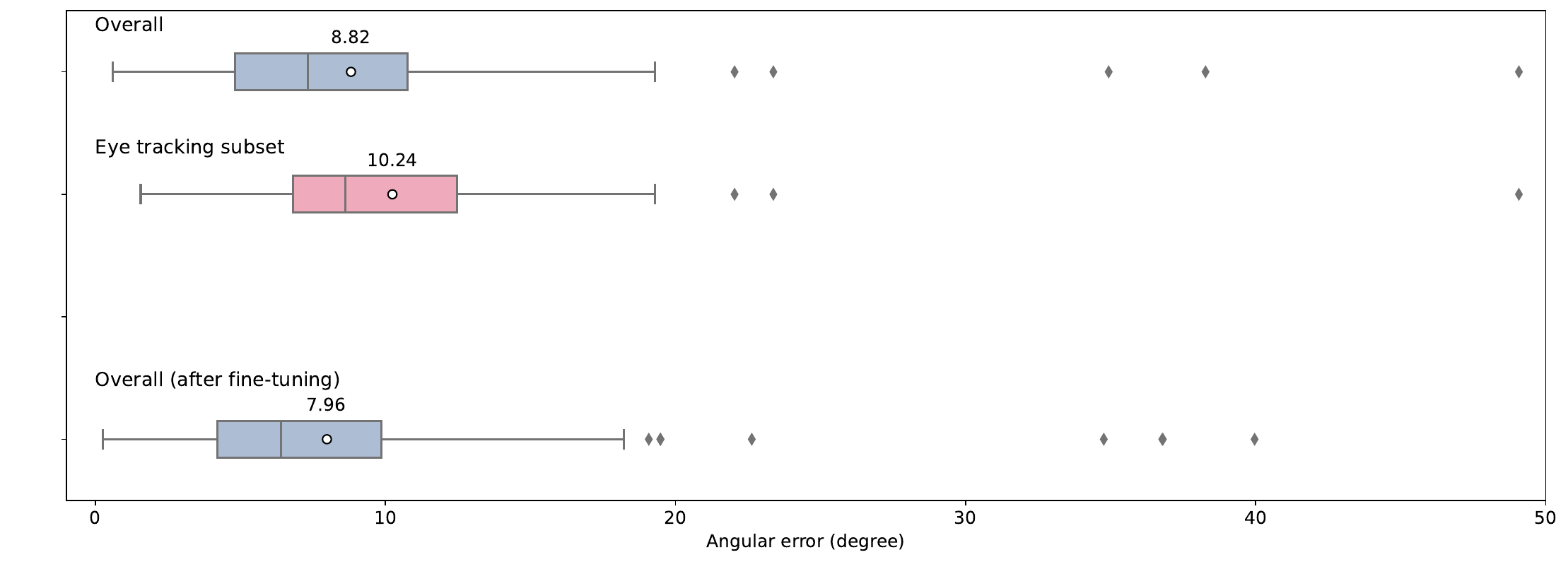}
    \caption{Angular error distributions of the pre-trained gaze estimator. From top to bottom, each box plot corresponds to the overall distribution of the pre-trained model, the distribution over the eye tracking subset, and the overall distribution of the fine-tuned model.\label{fig:wear_pupil_compare}}
    % \Description{Three box plots corresponding to different models and evaluation subsets.}
\end{figure}

Next, we investigate whether the data collected from our gamified system can contribute to evaluating and training the gaze estimation models. 
For this purpose, we trained the state-of-the-art Hybrid GazeTR model~\cite{cheng2022gaze} using the MPII-NV dataset. 
We use all 140 data points acquired in the gamified setting, which includes data from participants with and without the eye tracking headset, to test the pre-trained model.

Figure~\ref{fig:wear_pupil_compare} shows box plots illustrating the distribution of estimation errors of the pre-trained model. 
The ground-truth values are calculated based on the location of the letters, and the plots reflect the angular error between those true values and the estimated gaze vectors. 
The first plot corresponds to the error distribution of the pre-trained model.
The mean angular error was $8.82$ degrees, notably larger than expected in ordinary datasets. 
This indicates that our data presents quite a challenging scenario for gaze estimation. 
Furthermore, the second plot corresponds to the subset wearing the eye tracking headset, as used in Sec.~\ref{sec:eyetracker}. 
One should note that this data is typically more difficult due to the occlusion caused by the eye tracker itself. 
Even considering this factor, the error is much larger than in Fig.~\ref{fig:angular_comparison}, confirming the existence of room for improvement in estimation accuracy.

To further verify whether the model's accuracy improves with additional training using our data, we evaluated the precision of the model fine-tuned in a 3-fold cross-validation.
Specifically, we split the participants into three folds, where the data from the eye tracking subset were balanced in both fine-tuning and test sets.
We randomly chose 15 images from the fine-tuning group to train the model by ten epochs.
The third plot in Fig.~\ref{fig:wear_pupil_compare} corresponds to this fine-tuned evaluation, and the mean error is reduced from 8.82 to 7.96.
Please note that the number of images used in this experiment is much fewer than typically used for regular model training. 
However, the results show that even fine-tuning with a few images can enhance the model's overall accuracy, demonstrating the value of the collected images as training data.

\begin{figure}[t]
    \centering
    \includegraphics [width=0.8\linewidth]{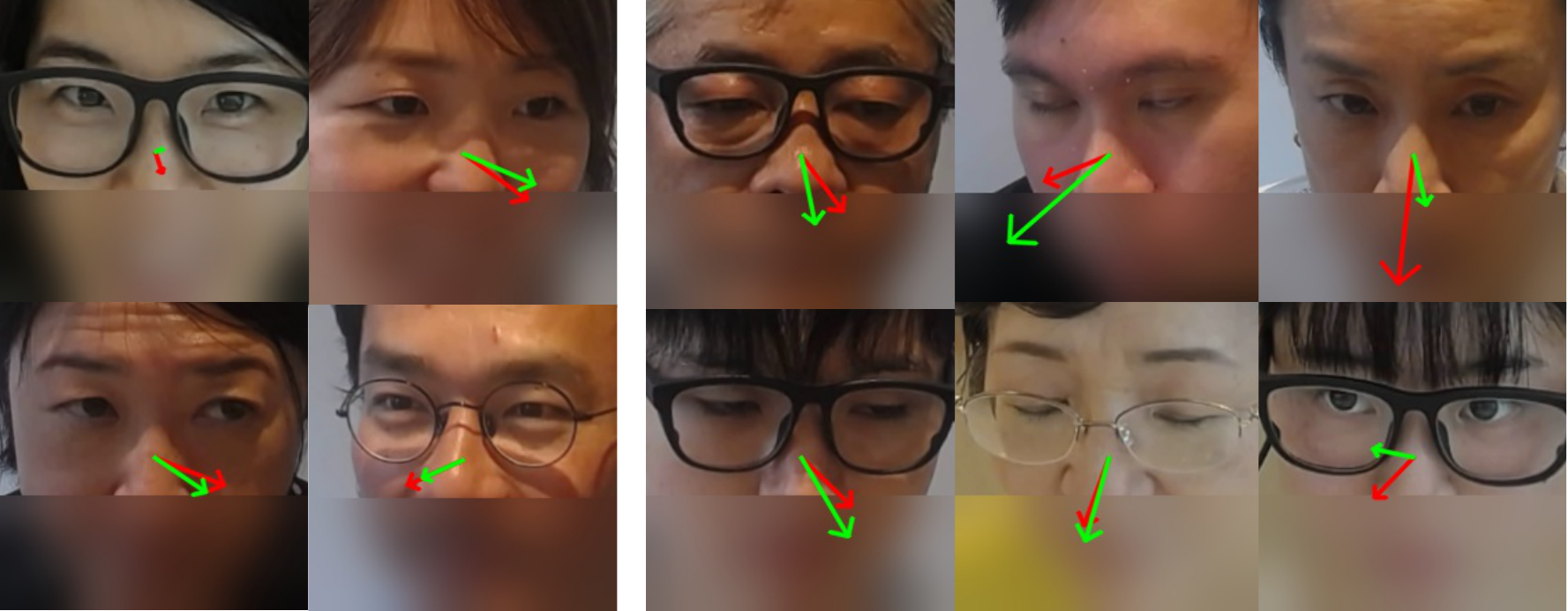}
    \caption{Examples of gaze estimation on our data. The red and green arrows represent the ground-truth and the predicted gaze directions, respectively.\label{fig:qualitative_gaze}}
    % \Description{Face images placed on tiles. Vectors indicating the direction of gaze are visualized for each.}
\end{figure}

Figure~\ref{fig:qualitative_gaze} presents examples of prediction results from the pre-trained gaze estimator. 
Red arrows display the ground-truth value reflecting the letter's location, while green arrows indicate the predicted gaze direction. 
Although we obtained permission to capture face images during our study, the lower portion of the face is intentionally obscured to prioritize privacy.
One interesting observation from the data obtained in this study is the diverse range of facial expressions exhibited by the participants. 
The images on the left show examples of such facial expressions, including laughing or engaging in conversation. 
These factors may have the potential to influence the accuracy of gaze estimation. 
The relationship between facial expressions and gaze estimation has not been thoroughly investigated, and our data collection methodology could shed light on this matter. 
Images on the right side represent a group of samples with higher error. 
This includes cases where estimation is fundamentally difficult, such as when a person's eyes are almost invisible, and cases close to outliers where the image was not taken correctly during the game.
The case where the participant blinks at the moment of capture is clearly undesirable, and it is important to design a process in which the players can verify the captured images more correctly. 

A significant correlation was observed between the error estimated by a pre-trained model and the error measured by an eye tracker ($r=0.72$, $p<0.01$). 
However, this correlation was largely due to the outlier data mentioned earlier. 
When this data was removed, no correlation was observed in the remaining data ($r=0.07$, $p=0.57$). 
The z-score of the outlier data calculated based on the error estimation by the pre-trained model was 6.02. 
This suggests that even a simple method could automatically eliminate this outlier. 
Among the data collected without using an eye tracker, there were two instances where the z-score exceeded 3.0, corresponding to the two rightmost images in the first row of Fig.~\ref{fig:qualitative_gaze}.

\subsection{User Experience from the Questionnaire}

\begin{figure}[t]
    \centering
    \includegraphics [width=1.0\linewidth]{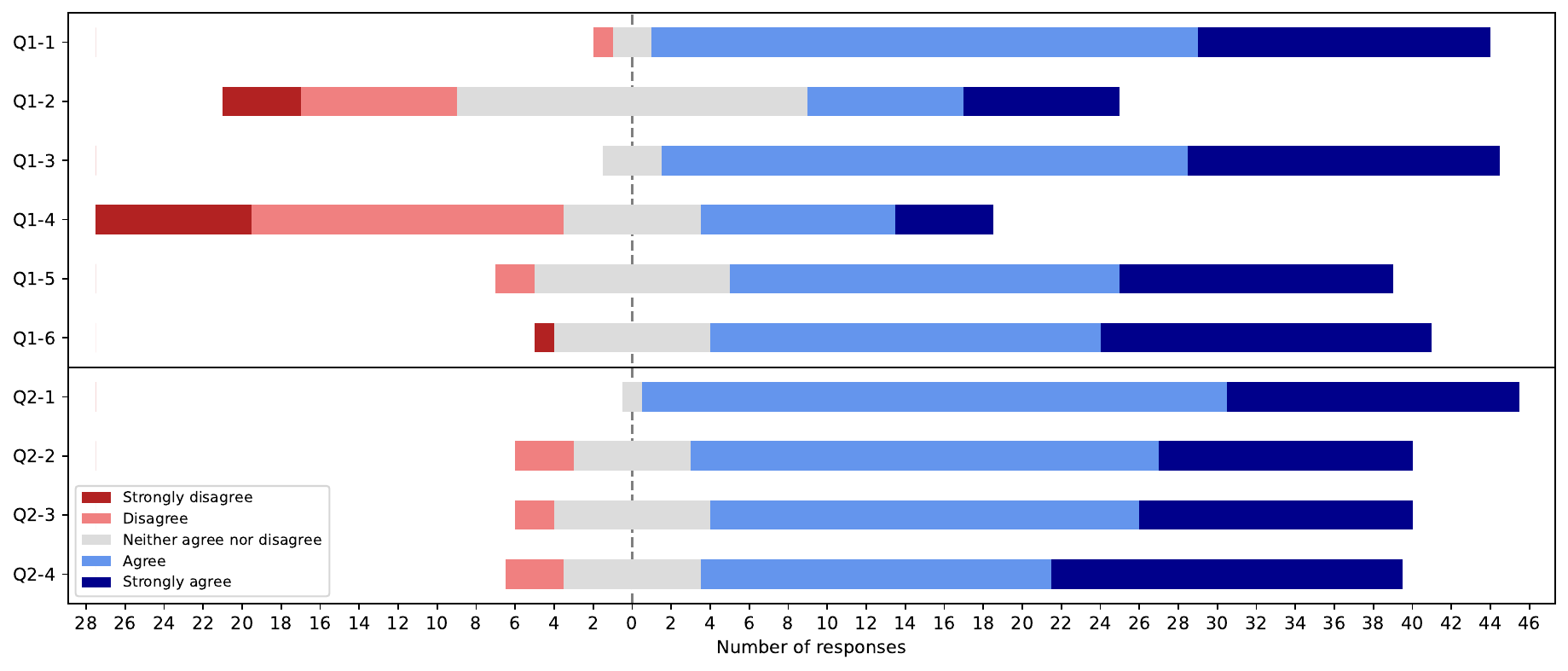}
    \caption{Responses from the questionnaire. Each chart corresponds to a question in Table~\ref{tab:qq}, as indicated on the horizontal axis. Different colors represent each option, from red for ``\textit{strongly disagree}'' to blue for ``\textit{strongly agree}''.\label{fig:response-likert}}
    % \Description{Ten stacked bar charts corresponding to each user question.}
\end{figure}

The five-point Likert scale ratings from the online questionnaire are shown as stacked bar charts in Fig.~\ref{fig:response-likert}.
The vertical axis shows the question IDs in Table~\ref{tab:qq}.
The horizontal axis shows the number of responses for each question.
Different colors represent each option, from red for ``\textit{strongly disagree}'' to blue for ``\textit{strongly agree}''.

The upper part (Q1) is aligned with questions about the game experience. 
Overall, we found favorable evaluations for the enjoyment of the experience, as observed in Q1-1, 3, and 5. 
Particularly, Q1-3, about how interesting it was to guess the words, did not receive any negative feedback. 
Furthermore, the question regarding the desire for repeated gameplay (Q1-6) received consistently positive responses, indicating high user satisfaction and replay value. 
Questions centered on the game's difficulty level (Q1-2, 4) revealed divided opinions, with positive and negative responses equally prevalent. 
The difficulty in gaze reading (Q1-2) showed a nearly symmetrical distribution from the respondents, whereas the challenge level of guessing the words (Q1-4) gathered several negative responses.
A significant medium positive relationship existed between Q1-2 and Q1-4 (Spearman’s rank correlation, $r=0.45$, $p<0.01$).
However, neither of their correlations with the question about interestingness (Q1-1, 3) was significant, and the difficulty did not necessarily mean they did not enjoy the game.

The lower section of the figure focuses on questions about the insights participants gained from the game experience. 
The predominantly positive responses suggest that the game can foster valuable learning experiences.
However, compared with Q2-1, a question asking if the game served as an opportunity to gain knowledge about eye tracking technology, we observed more negative responses for other questions related to a deeper understanding of the game's meanings and intentions. 

Out of all respondents, twelve individuals left comments in the open-ended section, most of which were brief due to the nature of the experimental setting. 
Five comments expressed gratitude for the insights gained from experience, such as ``\textit{I learned a lot}.'' 
Furthermore, three comments mentioned something about participating in the research, like ``\textit{I am glad to be part of the data analysis}.''
We also noted a range of impressions regarding difficulty level. 
A comment from P32 supplements the trends observed in Q1-2 and Q1-4; ``\textit{While I could guess the rough region, confirming the word from there was quite difficult}.'' 
This participant responded to Q1-2 as easy (``\textit{strongly agree}'') but Q1-4 as difficult (``\textit{disagree}'').
P30, in contrast, suggested that making the questions more challenging and complex would be acceptable, responding as easy (``\textit{strongly agree}'') to both Q1-2 and Q1-4.
A comment from P21 reflected the gap between the gaming experience and the real-world research community, stating that ``\textit{I do not mind my facial images being used by researchers I know. However, I would feel uncomfortable if they were used by a researcher unknown to me}.'' 
Finally, participant P45 speculated about potential applications of gaze-tracking technology; ``\textit{It would be amazing if feelings could be understood from one's gaze}.'' 
This indicates the participant's inspired anticipation towards further advancements and broader uses of gaze estimation.
\section{Discussions}

\subsection{Quality of the Collected Data}

\subsubsection{Label Accuracy}

The quantitative comparison of gaze label accuracy between the gamified and standard settings reveals several key insights. 
The wearable eye tracker experiments demonstrate that the gamified approach can acquire gaze labels with accuracy comparable to the standard data collection protocol, as evidenced by similar mean angular errors after removing the outlier in the gamified setting.
This finding suggests that the proposed gamified system is a viable alternative for collecting training data for appearance-based gaze estimation, especially when aiming to engage a diverse range of participants.

However, outliers in the gamified setting highlight the importance of implementing robust data filtering and quality control mechanisms. 
While the gamified approach can introduce outliers, it also captures more diverse and naturalistic face images, which are beneficial for robust model training.
The trade-off between data diversity and consistency should be carefully considered when designing gamified data collection systems. 
In designing a system, it is crucial to balance creating engaging experiences that capture lively face images and implementing appropriate data filtering and quality control measures.
Future iterations of the system should focus on minimizing the impact of outliers through improved user guidance, real-time feedback, and post-processing techniques.

As supplementary results, we confirmed that removing outliers via a pre-training model is possible to some extent. 
This direction can potentially become significant when dealing with large-scale data collection. 
However, it should be noted that automatically removing outliers can be challenging, as it may lead to too much data being lost.
Therefore, it is certainly desirable to be able to exclude during the game. 
Although a process for players to qualitatively review their captured images was implemented in our system, it was probably insufficient for them to notice label errors.
The development of real-time data quality feedback mechanisms could help mitigate the impact of outliers while maintaining an enjoyable user experience.

\subsubsection{Image Diversity}

The evaluation of the pre-trained gaze estimation model on the collected data provides valuable insights into the characteristics and potential utility of the gamified dataset.
The higher estimation errors compared to the model's performance on existing datasets suggest that the gamified data presents novel challenges and variability not encountered in standard benchmark datasets.
These errors were larger than those observed through the eye tracker experiment, suggesting a gap exists between the gaze estimation model's performance and upper-bound accuracy. 
This finding underscores the importance of collecting diverse and representative data that captures the complexity of real-world facial behavior.

Furthermore, the improvement in the model's performance after fine-tuning a subset of the gamified data demonstrates the potential of the collected dataset to enhance the generalization and robustness of gaze estimation models. 
The fine-tuning results indicate that the gamified data can be a valuable augmentation to existing training datasets.
However, it is important to note that the fine-tuning experiment was conducted on a relatively small scale, and further research is needed to assess the full potential of the gamified dataset for model training and evaluation. 
Future studies should also explore the impact of different training strategies, including domain adaptation approaches, and compare the performance gains across multiple gaze estimation architectures.

\subsection{User Experience}

Overall, reactions toward the gamified system were largely positive based on questionnaire responses. 
Participants indicated sustained interest in guessing words (Q1-3), enjoyable cooperation (Q1-5), and a desire to replay (Q1-6). 
While promising, it must be noted we cannot definitively state that our approach directly improved engagement without comparative data on the non-gamified setting and other possible design options. 
Still, willingness to repeat the game has promising implications for feasibility in, e.g., long-term public installations.

The participants responded positively to their interest in the technology. 
We did not initially anticipate the extent of their positive response to our inquiries about their grasp of the technology, but the overall response was favorable (Q2-1). 
Despite the need for further investigation to ascertain the accuracy of their comprehension, it seems we were somewhat successful in conveying the concept of data collection for appearance-based gaze estimation through the gaming experience. 
This makes a strong case for using gamified experiences to share complex technology concepts in a more accessible and engaging way.

Given the nature of the experiment, the participants' demographics were beyond our control, and their demographics can be seen as the potential audience for this game. 
In that sense, it is noteworthy that the age range of the participants in this experiment was broadly distributed from 18 to 79. 
Stimulating the interest and enjoyment of a diverse group of participants is a significant challenge from both data collection and technical communication perspectives. 
Our physical game design has been carried out with the acquisition of such a diverse range of players in mind, and the participants' age distribution and their survey results demonstrate a glimpse of its potential.

As mentioned at the beginning, privacy is a significant issue when dealing with face images, and it must be emphasized again that this study's approach does not directly address this problem. 
Despite clear explanations and obtained consent from game participants in our study, a few still expressed concerns about their face images being used in research (Q2-4). 
This aspect becomes particularly critical when the game is set up as a public installation. 
It is highly possible that merely displaying text explanations is insufficient to ensure players fully understand the research context and data usage. 
A more in-depth experience design will be necessary for future work, moving beyond simply acquiring superficial engagement.

\subsection{Limitations and Future Work}

\subsubsection{Study Conditions}

While including a standard lab-based data collection setting provides a baseline for label accuracy, it is indeed insufficient to evaluate user experience. 
Ideally, data collection without game elements should be conducted with visitors in public spaces to demonstrate the effects of gamification. 
While it is unlikely that standard data collection protocol is enjoyable for participants, it cannot be denied that those who find the game particularly difficult might view regular data collection methods more favorably.
Moreover, manual administration by researchers in the gamified setting may inadvertently influence user behavior and experience, and a fully autonomous system could reveal different characteristics. 
Conducting rigorous comparative experiments in the future would be necessary to demonstrate the significance of gamification from the user's perspective.

In addition, we have not conducted a detailed comparison to test the effects of our design choices, such as cooperative game design and the introduction of physical elements. 
It must be noted that implementing an online version of this game is not trivial, considering the difficulty of interpreting and associating the gaze direction of a remote partner through a webcam.
However, a fully software-based implementation could be highly beneficial from the data collection perspective.

Regarding the survey design, our current questionnaire focuses primarily on the participants' enjoyment and perception of the gamified experience. 
While these factors are essential for assessing user engagement, the survey could be expanded to include more specific questions related to the data collection process and the quality of the collected data. 
Including common measures such as IMI~\cite{deci1985intrinsic} and PXI~\cite{abeele2020development}, a detailed analysis of the gaming experience is also an important issue for future research.

\subsubsection{Game Design}

Our gamified procedure takes more time to capture a single image than standard data collection methods. 
As highlighted in previous research~\cite{bragg2021asl}, this can be viewed as a somewhat unavoidable outcome of gamification.
As discussed above, designing game experiences that can withstand long-term public installations is an important challenge to overcome this shortcoming. 
By carefully addressing privacy concerns, it will be necessary to actively consider using the gamified system through exhibitions and other ways of presentation.
For example, by introducing a visualization of how the performance of the gaze estimation model has improved with players' participation so far, ways to enhance engagement from a more long-term perspective could be devised.

Considering the wide variation in difficulty evaluation among players, it is important to introduce a mechanism that adaptively changes the difficulty level. 
In this game, it could be possible to address this by varying the number of hint letters based on the required time and the correctness of the previous answer. 
As a more advanced approach, eye movements estimated by the appearance-based method might help predict the player's cognitive load. 
Incorporating eye tracking data into the gaming experience in various ways also holds significant potential from a technical communication standpoint.

Another design challenge involves the inherent bias in each character's appearance frequency when using regular words. 
An approach might be distributing frequently appearing characters evenly, but this could make it harder for players to find characters. 
Another aspect that warrants further investigation is the impact of the reversed letter layout. 
In the current implementation, the letters on the transparent board are arranged in a mirrored layout for one of the players. 
While this design choice ensures that both players have the same physical reference for each letter, it may affect the readability and search efficiency of the player viewing the mirrored layout. 
Future studies could explore alternative letter arrangements, such as using a transparent display that dynamically adapts the letter orientation based on the player's perspective. 
Additionally, collecting subjective feedback from participants regarding the ease of finding and focusing on specific letters could provide valuable insights into the impact of the letter-layout on the gameplay experience.
Such design considerations are also crucial when developing the game in other languages, such as English, which is an important future task for us.

\subsubsection{Technical Understanding and Communication}

Although user surveys resulted in positive feedback, one critical task is to foster the right understanding of ML and gaze estimation through the game experience. 
Explaining the concept of appearance-based gaze estimation to non-expert audiences is not always easy. 
Consequently, grasping how the data collected from this game aids in model training can be challenging. 
Specifically, when used in long-term installations, it is essential to clearly show players how the performance of the gaze estimation model could be improved due to data collected from past game experiences. 
This is a key area we are considering for future developments. 
It is crucial to allow players to see the concrete results of their participation and their data's contribution to the model training.

Moreover, it is essential to consider using this game to stimulate participant discussions and increase their engagement with the technology, especially when paired with activities like workshops. 
The short duration of the game experience alone may not sufficiently cover deep discussions about ethical issues related to facial images and gaze estimation. 
However, the potential benefit of fostering positive impressions through these game experiences could warrant continued exploration in the future. 
We believe the strategy of encouraging understanding and conversation about ML technology by turning the data collection process into a game could extend beyond gaze estimation. 
This approach might also promote technical literacy in other AI fields by providing an enjoyable platform for dialogue and learning.

\section{Conclusion}

This research proposed a novel method to transform the data collection process for appearance-based gaze estimation into an engaging game for the general public. 
The game, uniquely complemented by a physical device, is designed for two seated players who communicate words using their gaze through a transparent letter board. 
In this scenario, the facial images of the player conveying the words can be recorded. 
This allows the collection of training data, paired with the gaze position, for gaze estimation. 
The system is intended to stimulate technological interest in as many individuals as possible and is designed to enable people to comprehend it accurately and to participate in data collection while having fun. 
This depicts a blending of interactive engagement and scientific data collection, opening new possibilities for implementing data collection approaches in applied ML research.

We validated the effectiveness of the proposed system by installing it in a space accessible to the public. 
The data accuracy was evaluated using the gaze positions obtained from a wearable eye tracker. 
While the accuracy was comparable to standard data collection, it was observed that extreme outliers could occur. 
The accuracy of gazes predicted by the pre-trained gaze estimation model was inferior to that on standard datasets, suggesting that the data collected with our system is valuable for the evaluation and training of gaze estimation models. 
The feedback collected from participants' surveys was generally positive, confirming that we achieved our goal of engaging people in data collection while fostering interest in the technology. 
Looking forward, we plan to explore other usage scenarios, including long-term exhibitions or pairing with workshops, and consider expanding our approach beyond gaze estimation.

\section*{Acknowledgements}
This work was partly supported by the UTokyo Ushioda Foundation, the Foundation for the Promotion of Industrial Science, and JST CREST Grant Number JPMJCR19F2, Japan. 

\section*{Declaration of Interest Statement}
The authors have no competing interests to disclose.

\bibliographystyle{apacite}
\bibliography{main}

\end{document}